\documentclass[10pt]{article}
\usepackage{multicol}
\usepackage[latin9]{inputenc}
\usepackage[T1]{fontenc}
\usepackage{longtable}
\usepackage[a4paper, total={7in, 10in}]{geometry}
\usepackage{float}
\usepackage{graphicx}
\usepackage{epstopdf}
\usepackage{subfigure}
\usepackage{amssymb}
\usepackage{fancybox}
\usepackage{amsmath}
\usepackage{epsfig}
\usepackage{dsfont}
\usepackage{fancyhdr}
\usepackage{picins}
\begin{document}
\include{00README.XXX}

\begin{center}
\begin{Large}
The extraction of $^{229}$Th$^{3+}$ from a buffer-gas stopping cell \end{Large}
\end{center}

\begin{center}
\begin{large}
L. v.d.Wense$^1$, B. Seiferle$^1$, M. Laatiaoui$^{2,3}$, and P.G. Thirolf$^1$ 
\end{large}
\end{center}
$^1$ Ludwig-Maximilians-Universitõt M³nchen, Am Coulombwall 1, Garching, Germany\\
$^2$ GSI Helmholtzzentrum f³r Schwerionenforschung GmbH, Planckstr. 1, Darmstadt, Germany\\
$^3$ Helmholtz Institut Mainz, Johann-Joachim-Becherweg 36, Mainz, Germany\\[0.3cm]
In the whole landscape of atomic nuclei, $^{229}$Th is currently the only known nucleus which could allow for the development of a nuclear-based frequency standard, as it possesses an isomeric state of just 7.6 eV energy above the ground state. The 3+ charge state is of special importance in this context, as Th$^{3+}$ allows for a simple laser-cooling scheme. Here we emphasize the direct extraction of triply-charged $^{229}$Th from a buffer-gas stopping cell. This finding will not only simplify any future approach of $^{229}$Th ion cooling, but is also used for thorium-beam purification and in this way provides a powerful tool for the direct identification of the $^{229}$Th isomer to ground state nuclear transition. 

\begin{multicols*}{2}
\section{Introduction}
$^{229}$Th was first considered to possess a low energy isomeric state in 1976 by Kroger and Reich, where the enery was estimated to be below 100 eV \cite{Kroger_Reich}. Later measurements indicated an energy as low as 3.5 eV \cite{Helmer_Reich2}, which was the most accepted value until 2007, when it was corrected to 7.6 eV (corresponding to a wavelength of about 160 nm) by applying a high resolution cryogenic $\gamma$-ray detector \cite{Beck1}. This is an unusally small energy for a nuclear state, which even matches energies typically involved in atomic shell processes and, as a unique feature, conceptually allows to drive the corresponding isomeric ground-state transition using laser technology \cite{Matinyan}. This in turn has led to the proposal of interesting applications, e.g. to use the corresponding transition as a highly stable nuclear-based frequency standard \cite{Peik1}. Besides its small energy, relatively long half-lives of up to several hours (depending on the actual energy, the electronic environment and the correspondingly allowed decay channels) have been predicted \cite{Karpeshin1}, which would result in an extremely small natural linewidth of $\Delta E/E\approx 10^{-20}$. Combined with an expected high resistance against external influences like electric and magnetic fields, the nuclear based clock could potentially outperform the existing atomic clock technology \cite{Peik1}.\\
However, despite significant experimental effort, up to today no direct unambiguous identification of the isomeric decay was possible and also the properties of the isomeric transition, namely energy, half-life and photonic branching ratio, are not known precisely enough to allow for a direct optical access of the isomer to ground-state nuclear transition. Our experiment is aiming for a direct detection of the isomeric decay, which will then allow for a determination of the decay parameters and therefore pave the way for the development of a nuclear-based frequency standard.

\begin{figure*}
 \resizebox{1.0\textwidth}{!}{%
 \includegraphics{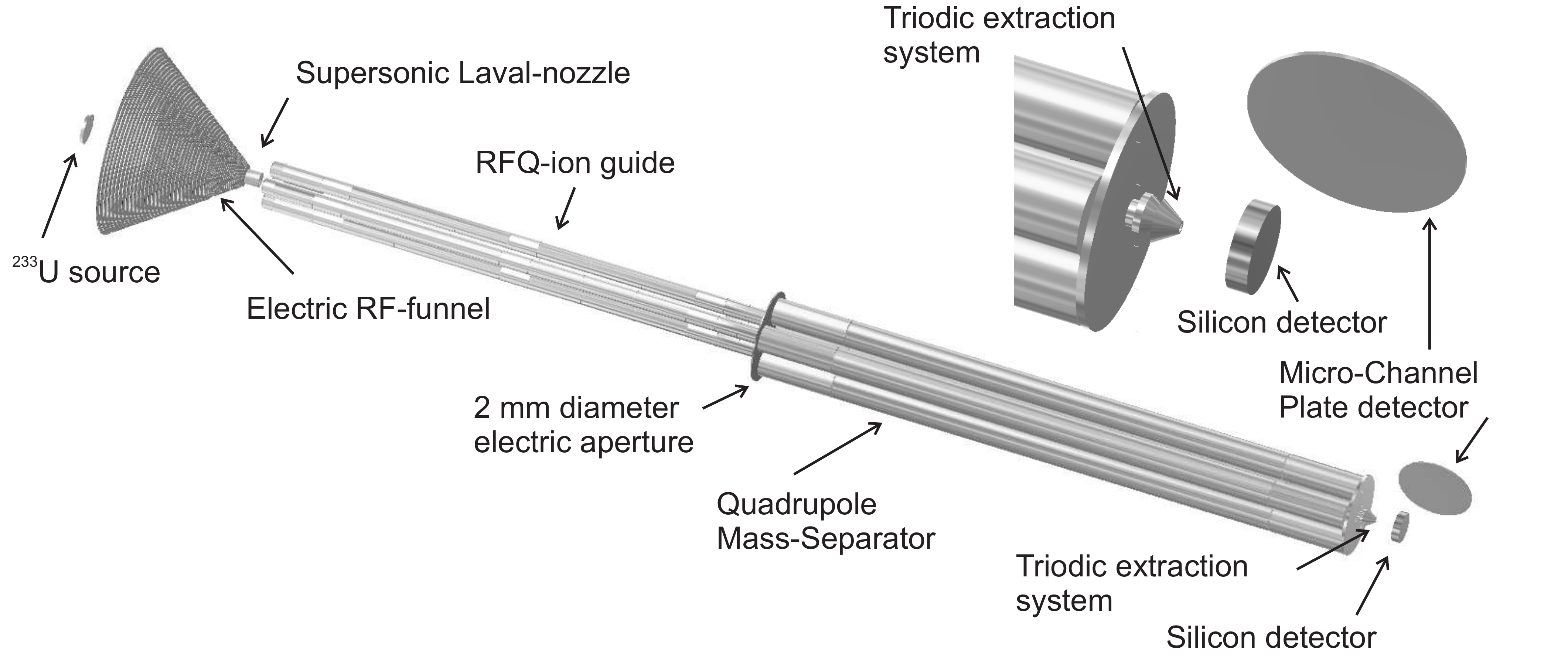}}
 \caption{Overview over the ion extraction system applied for the production of a highly purified $^{229}$Th ion beam. Also the detectors used for beam diagnostics are shown (details see text). The overall length of the system is about 80 cm.}
 \label{setup}
\end{figure*}

\section{The ion extraction system}
\begin{figure}[H]
\begin{center}
 \includegraphics[totalheight=6cm]{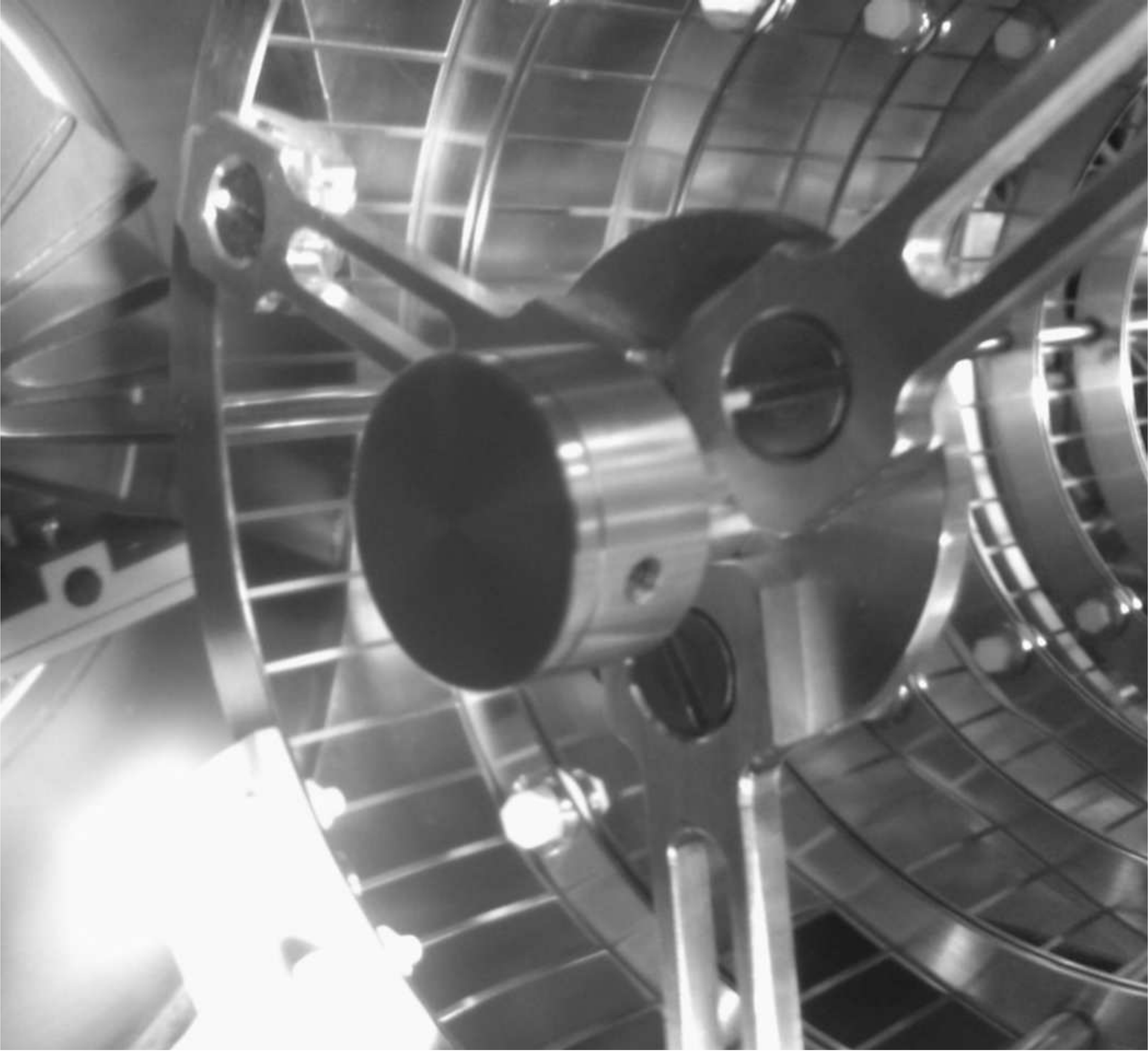}
 \caption{$^{233}$U source of 20 mm diameter mounted into the buffer-gas stopping cell in front of the RF funnel (not visible to the left of the source).}
 \label{source}
 \end{center}
\end{figure}

A conceptual overview over the ion extraction and purification system is shown in Fig. \ref{setup}. Within this experimental approach, $^{229m}$Th is populated via a 2\% decay branch in the $\alpha$ decay of $^{233}$U \cite{Wense1}. For this purpose, an $^{233}$U (UF$_4$) source of about 200 kBq activity, evaporated onto a 20 mm diameter stainless steel plate, is used and mounted into a buffer-gas stopping cell \cite{Neumayr1}. A picture of the source, when mounted into the buffer-gas stopping cell, is shown in Fig. \ref{source}. Besides $^{233}$U, also $^{232}$U is contained as a trace contamination of about $3.9\cdot10^{-7}$ in the source material due to the radiochemical production process.\\
A fraction of the $^{229}$Th $\alpha$-recoil ions, as produced in the $\alpha$ decay of $^{233}$U, is leaving the source material due to the kinetic recoil energy of 84.3 keV.

\noindent All $\alpha$-recoil ions leaving the source are stopped within 40 mbar helium with a purity level of 99.9999~\%, which is further cleaned by catalytic purification, a cryo trap and a getter pump to the ppb level. The buffer-gas stopping cell itself is strictly designed to ultra-high vacuum standards and bakeable up to 180░C. Besides the uranium source, the buffer-gas stopping cell also contains an electric RF(+DC)-funnel system, consisting of 50 ring electrodes converging in diameter towards the exit of the stopping cell. RF voltages with an amplitude of 220 V$_{\text{pp}}$ at 850 kHz, alternating in phase by 180░ between two electrodes, are applied to the funnel system, resulting in a repelling force, which prevents the ions from charge capture at the chamber walls. Further, a DC voltage gradient of 4 V/cm is applied in order to drag the ions through the gas background towards the chamber exit. The chamber exit is formed by a Laval nozzle with a 0.6 mm diameter nozzle throat. During extraction, the helium forms a supersonic gas jet when entering a second vacuum chamber with a typical background pressure of $2\cdot 10^{-2}$ mbar. The ions are following the gas flow and are successively injected into a radio frequency quadrupole (RFQ) ion guide system. The RFQ consist of four rods of 11 mm diameter and 10 mm inner rod distance. It is segmented into 12 parts, to each of which a separate DC voltage is applied in order to allow for a voltage gradient, which drags the ions through the remaining helium background. The applied RF amplitude is 200 V$_{\text{pp}}$ at 880 kHz. At the RFQ exit a sub-mm diameter low energy ion beam has formed due to phase-space cooling by the buffer-gas background.\\
At this point still all $\alpha$-recoil ions of the source in use are contained, especially also isotopes of the $^{233(232)}$U decay chains below $^{229(228)}$Th. As some of them are short-lived $\alpha$ and $\beta^-$ emitters, they are a potenial source of background, which could dilute any obtained signal of the $^{229m}$Th isomeric decay. In order to exclude this source of background, a customized quadrupole mass separator (QMS) was constructed, which allows for complete separation of $^{229}$Th from its daughters in the 1+, 2+ and 3+ charge states at a high transmission rate (a mass resolving power of $m/\Delta m\approx 150$ at about 70\% transmission rate was achieved). The QMS has a length of 30 cm with further 5 cm Brubaker lenses at entrance and exit. The rods have a diameter of 18 mm and an inner diagonal distance of 15.96 mm to each other (the design values were taken from \cite{Haettner}). For extracting $^{229}$Th$^{3+}$ it was operated at an RF amplitude of 600.5 V$_{\text{pp}}$ at 925 kHz. The DC voltage was set to 50.15 V.\\
After mass separation, the ions are extracted by a triodic extraction system\cite{Wense1} from the QMS and the isomeric decay is expected to take place. Due to its low energy, the isomer in the neutral thorium atom is expected to decay predominantly by internal conversion (IC). An internal conversion factor as large as $10^{9}$ has been predicted, leading to a half-life as short as $10^{-5}$ s \cite{Karpeshin1}. However, if the ions are kept charged, the IC decay channel is energetically forbidden, as the 2+ ionization potential of thorium is 11.9 eV and thus exceeds the predicted isomeric energy of 7.6 eV. In this case, the photonic decay channel could become dominant and long half-lives of up to $10^{4}$ s have been predicted. Besides the common photonic and electronic decay channels, the exceptionally low energy of $^{229m}$Th may also lead to a stronger coupling of the isomer to its electronic environment and bound internal conversion (BIC) \cite{Karpeshin1} effects as well as phononic coupling might play a non-negligible role for its decay.\\

\begin{figure*}
 \resizebox{1.0\textwidth}{!}{%
 \includegraphics{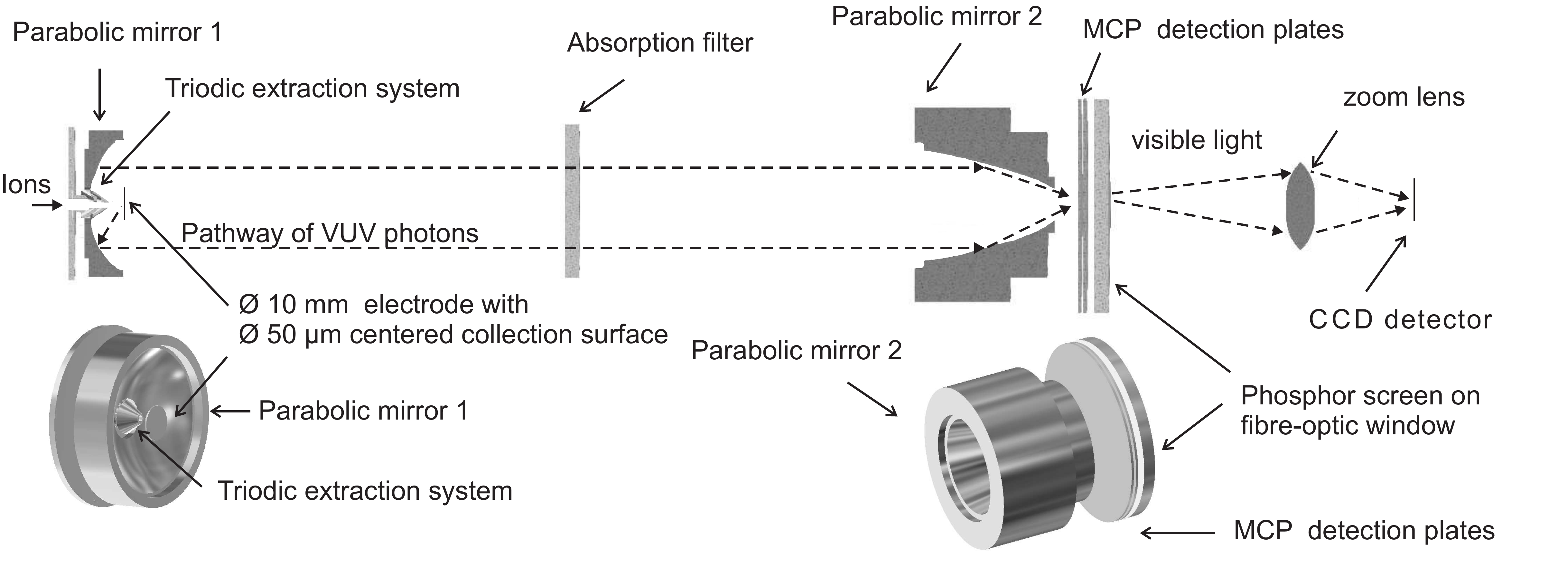}}
 \caption{Overview over the optical setup developed for the search of a photonic decay signal of the $^{229m}$Th ground-state transition (upper row). The bottom row displays two detailed views of the annular parabolic mirrors and the corresponding extraction/collection and detection systems.}
 \label{opticsetup}
\end{figure*}
\section{The optical system}
The described experiment allows to probe for a photonic as well as an electronic (IC) decay channel. In order to probe for an IC decay channel, the extracted ions can directly be accumulated with low kinetic energy onto the surface of an MCP detector, which allows for a spatially resolved signal read out. However, the search for a photonic decay channel requires a more sophisticated setup\cite{Wense1,Seiferle} as shown in Fig. \ref{opticsetup}. The thorium ions, leaving the triodic extraction system, are collected onto a 50 $\mu$m diameter micro electrode, which lies in the center of a 10 mm diameter circular electrode. While the electrodes are copper-based, a MgF$_2$ coating is foreseen in order to prevent the ions from direct discharge at the electrode surface.\\
Allowing the ions to stay charged opens the way for the search of a photonic decay channel. The long-lived isomeric decay is expected to take place on the collection surface, which itself is positioned in the focus of an annular parabolic mirror of 40 mm aperture and 10 mm focal length. MgF$_2$ coated aluminum acts as the mirror surface and offers a relatively high reflectivity of $\ge$70 \% in the vacuum ultra-violet region around 160 nm. A second annular parabolic mirror focuses the collimated light onto the detection plane. This focusing mirror is extremely deep with a focal length of 2 mm at 40 mm aperture. Thus its focal spot lies behind the mirror plane due to a 12 mm diameter central hole and in this way keeps the image distance as short as possible, thus leading to small image magnification and a good signal contrast. The simulated focal spot size under optimum alignment conditions is 70 $\mu$m. The advantage of this parabolic-mirror based all-reflective optical system is that it does not suffer from any spherical or any chromatic aberration. Further, it still offers a significant reflectivity at small wavelengths of about $\ge$130 nm. These are very promising features, having in mind that the actual energy of the isomeric transition is still suffering from big uncertainties \cite{Sakharov}.\\
The region of collimated light between the two mirrors allows to apply turnable absorption filters with a sharp absoption edge for the wavelength determination of detected light. A micro-channel plate (MCP) detector, combined with a phosphor screen, acts as a position sensitive single photon detector with 100 $\mu$m resolution.  The detector is of chevron geometry with 75 mm diameter, the front plate is CsI-coated in order to enhance the quantum efficiency in the VUV region. Typically, quantum efficiencies of about 10 \% are achievable at around 160 nm wavelength. The phosphor screen is monitored by a CCD camera for position sensitive read-out.\\
\section{The extraction of $^{229}$Th$^{3+}$}
\begin{figure*}
 \resizebox{1.0\textwidth}{!}{%
 \includegraphics{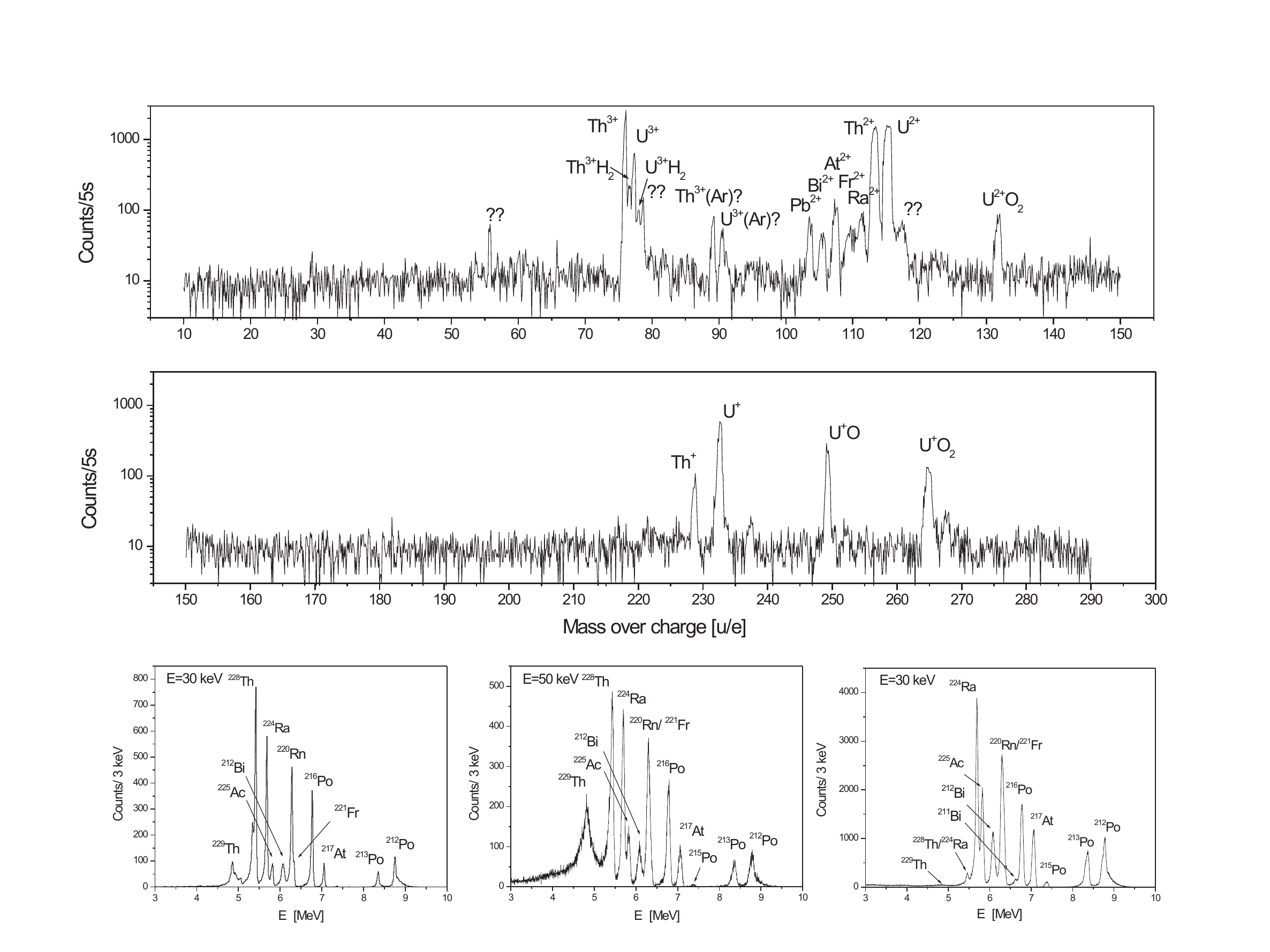}}
 \caption{Upper panel: mass scan of the extracted recoil ions from 10 to 290 u/e. The 1+, 2+ and 3+ charged species are clearly visible. Lower panel: $\alpha$ energy spectra used for the determination of the extraction efficiency of $^{229}$Th$^{3+}$ from the buffer-gas stopping cell. The left spectrum ($\Delta$ E=30 keV) corresponds to 100 days of detection after 5 days of extraction of Th$^{3+}$ onto the surface of a silicon detector. The right spectrum ($\Delta$ E=50 keV) results from 32 days of measurement, started after 5 days of direct recoil ion implantation into the silicon detector surface and further 200 days of decay. The long decay time was chosen, as directly after implantation, the $\alpha$ spectrum was heavily dominated by the short-lived daughter isotopes of the $^{229,228}$Th decay chains, as can be seen from the inset spectrum that shows the $\alpha$ decays as obtained after 1 day of decay and 1 day of detection.}
 \label{mass_scanfig}
\end{figure*}

A large absolute number of extracted thorium ions is a major prerequisite in view of the search for any isomeric decay signal. In order to determine the extracted absolute amount of thorium $\alpha$-recoil ions for the different charge states, a complete mass scan was performed using an MCP detector (Hamamatsu, type F2223) placed in about 5 mm distance behind the triodic extraction system and operated in single-ion counting mode with -2000 V surface voltage \cite{Wense2}. The complete scan is shown in Fig. \ref{mass_scanfig}. The scan reveals significant thorium extractions in the 2+ and 3+ charge states. Only a minor amount of recoil ions populates the 1+ charge state. Besides thorium, also uranium is extracted in all three charge states. The complete $^{233}$U decay chain is visible in the 2+ charge state. Please note, that the daughter isotopes of $^{229}$Th are not extracted with the same intensity as $^{229}$Th, as the decay chain is not in equilibrium. Assuming a detection efficiency of the MCP detector of about 50 \% (as being a typical value for heavy ions and later confirmed by silicon detector based measurements), the absolute thorium extraction rates are obtained to be $\sim 10^{3}$/s for Th$^{3+}$, $ \sim 550$/s for Th$^{2+}$ and $\sim 35$/s for Th$^{1+}$, respectively.\\
A silicon detector measurement was performed, aiming for a direct detection of the $^{229}$Th $\alpha$ decay. The Si detector (Ametek, type BU-016-300-100) was placed in a distance of about 5 mm behind the exit of the triodic extraction system. As the half-life of $^{229}$Th is 7932 years, a long continuous extraction time of 5 days was chosen in order to accumulate enough $^{229}$Th to make the $\alpha$ decays visible. During that time, the $^{229}$Th$^{3+}$ ions were directly accumulated onto the surface of the Si detector, which was set to an offset voltage of -1250 V in order to fully attract all extracted ions (no bias voltage was chosen, as the decay detection was carried out later). An MCP detector was operated in a 90 degree off-axis position (see Fig. \ref{setup}) in order to allow for a QMS tuning by direct ion counting prior to any accumulation on the Si detector. 
\end{multicols*}
\begin{multicols*}{2}
\noindent After the accumulation, the silicon detector was removed to a different vacuum chamber and the occuring $\alpha$ decays were detected for 100 days. The obtained $\alpha$-energy spectrum ($\sim$30 keV resolution) is shown on the left-hand side of Fig. \ref{mass_scanfig} (lower panel). As can be inferred, besides the decay chain of $^{229}$Th, the decay chain of $^{228}$Th is also contained in the spectrum, as the QMS mass-resolving power is not sufficient to separate $^{229}$Th and $^{228}$Th in the 3+ charge state. The obtained decay rate for $^{229}$Th is 54$\pm$5 per day, which corresponds to an accumulation rate of $(1.0\pm0.1)\cdot 10^{3}$ s$^{-1}$. This measurement validates the high thorium extraction rate in the 3+ charge state.\\
In order to also estimate the combined extraction and purification efficiency of the setup, a third measurement was carried out to determine the total emission rate of $^{229}$Th $\alpha$-recoil ions of the $^{233}$U source. For this purpose, all $\alpha$-recoil ions emitted by the source were directly implanted into the surface of a silicon detector (Ametek, type BU-017-450-100) placed in 5 mm distance in front of the source. After this implantation, the short-lived daughter isotopes were heavily dominating the $\alpha$-energy spectrum, which did not allow for the direct determination of the $^{229}$Th decay rate in the first place (Fig. \ref{mass_scanfig} lower right panel). However, after 200 days of decay, the daughter activity (dominated by the decays of $^{225}$Ra (T$_{1/2}$=14.9 d) and $^{225}$Ac (T$_{1/2}$=10.0 d), respectively) had faded away and the $^{229}$Th activity became visible. The corresponding $\alpha$-energy spectrum after 32 days of detection is shown in the center of Fig. \ref{mass_scanfig} (lower panel). The $^{229}$Th detection rate is found to be $4.3\cdot10^{-3}$ s$^{-1}$, leading to a total number of $3.1\cdot10^{9}$ implanted $^{229}$Th recoil ions. Taking into account that $\sim$67 \% of the recoil ions emitted into the hemisphere were implanted into the silicon detector, the rate of $^{229}$Th $\alpha$-recoil ions emitted from the source is estimated to be $10.7\cdot10^3$ s$^{-1}$. This is by a factor of 2.8 enhanced, compared to the expectations based on TRIM simulations. The discrepancy might be caused by the special production process of the $^{233}$U source via the evaporation technique. This technique may have led partly to crystal formation opening the possibility for channelling, which is not considered in the TRIM simulations. Based on these findings, the extraction and purification efficiency for the $^{229}$Th $\alpha$-recoil ions is estimated to be about 10 \% in the 3+ charge state, while 5.5 \% and 0.34 \% are inferred for the 2+ and 1+ charge states, respectively.

\section{Summary and perspectives}
An extraction and purification efficiency of 10 \% for $^{229}$Th$^{3+}$ from a buffer-gas stopping cell was obtained. To our knowledge, this is the first time that a significant extraction rate for triply-charged ions from a buffer-gas stopping cell was reported. Besides the ultra-high cleanliness of the extraction system in use, the reason for this finding lies in the exceptionally small 3+ ionization potential of Th of just 18.3 eV. During stopping and extraction of highly charged ions in a He environment, charge capture of the ions takes place until it is energetically favourable for the electron to stay attached to the helium atom rather than leading to a further charge-state reduction of the stopped ions. The energetical limit in this case is 24.6 eV, which is the ionization potential of He. Most 3+ ionization potentials are above this value and are therefore further reduced. This, however, is not the case for thorium, which preserves the 3+ charge state.\\
As a lucky coincidence, within the whole decay chains of $^{233}$U and $^{232}$U, respectively, only thorium allows for an extraction in the 3+ charge state. This leads to the possibility to use already the charge state separation as a filter to purify the thorium ions from their short-lived daughter isotopes and therefore provides a useful tool in the search for any $^{229}$Th isomeric decay. Experimentally, all short-lived daughter nuclei are found to be suppressed in the 3+ charge state by a factor of $10^{-3}$ to $\le10^{-4}$ compared to their 2+ extraction rates \cite{Wense2}. Further, the 3+ charge state of thorium has the advantage that it allows for a direct laser-cooling scheme \cite{Campbell1}. Thus the direct extraction of $^{229}$Th$^{3+}$ will significantly simplify any approach of laser cooling, which is required for the development of a $^{229}$Th-ion based nuclear clock, e.g. by employing laser-excitation on cold stored $^{229m}$Th$^{3+}$ ions in a Paul trap\cite{Peik1}.\\
The absolute observed extraction rate of about $10^{3}$ ions per second (for the 200 kBq $^{233}$U source) allows for an estimation of the achievable signal intensity for the detection of any photonic isomeric decay. Assuming 2\% of the $^{229}$Th $\alpha$-recoil ions to be in the isomeric state, further losses of intensity due to efficiencies of the ion collection, the optical system and the MCP detection expected to be 40 \%, 20\% and 10 \%, respectively, the detected decay rate could be 0.16 s$^{-1}$. Measurements revealed an achievable optical spot-size of about 100 $\mu$m diameter and the MCP dark count rate was measured to be 0.01 cts/s mm$^2$. This leaves us with a signal contrast of $\sim 2.0\cdot 10^{3}$, provided that no radiationless losses occur. The achievable signal contrast could be further increased when applying a $^{233}$U source of higher $\alpha$-recoil ion activity.\\[0.5cm]

\begin{Large}
\begin{bf}
Acknowledgements\\[0.2cm]
\end{bf}
\end{Large}
This work was supported by DFG (Th956/3-1) and by European Union's Horizon 2020 research and innovation programme under grant agreement No 664732 "nuClock".

\end{multicols*}
\end{document}